\documentclass[conference]{IEEEtran}

\usepackage[T1]{fontenc}
\usepackage{lmodern}

\usepackage[cmex10]{amsmath}
\usepackage{amssymb}
\usepackage{bm}
\usepackage{mathtools}

\usepackage[ruled,vlined]{algorithm2e} 

\usepackage{algorithmic}
\usepackage{float}
\usepackage{placeins}  
\usepackage{graphicx}
\usepackage{pgfplots}
\pgfplotsset{compat=1.17}  
\usepackage{tikz}
\usetikzlibrary{decorations.markings,decorations.pathmorphing}

\usepackage{subcaption}

\usepackage{booktabs}
\usepackage{multirow}
\usepackage{array}
\usepackage{tabularx}

\usepackage{cite}
\usepackage{xcolor}
\usepackage{textcomp}
\usepackage{gensymb}
\usepackage{siunitx}
\usepackage{enumitem}
\usepackage{afterpage}
\usepackage{amsthm}

\allowdisplaybreaks

\begin{document}
%
\title{Physics-Aware RIS Codebook Compilation for Near-Field Beam Focusing under Mutual Coupling and Specular Reflections}
\author{\IEEEauthorblockN{
Alexandros I. Papadopoulos\IEEEauthorrefmark{1}\IEEEauthorrefmark{2}, Maria Anna Pistela\IEEEauthorrefmark{1},
Dimitrios Tyrovolas\IEEEauthorrefmark{3}\IEEEauthorrefmark{4},
Antonios Lalas\IEEEauthorrefmark{1}, \\
Konstantinos Votis\IEEEauthorrefmark{1}, 
Sotiris Ioannidis\IEEEauthorrefmark{4}\IEEEauthorrefmark{5},
George K. Karagiannidis,~\IEEEmembership{Fellow,~IEEE}\IEEEauthorrefmark{3},
Christos Liaskos\IEEEauthorrefmark{2}
}                                     
\IEEEauthorblockA{\IEEEauthorrefmark{1}
 Information Technologies Institute, Centre for Research and Technology Hellas (CERTH), Greece} 
\IEEEauthorblockA{\IEEEauthorrefmark{2}
Computer Science and Engineering Department, University of Ioannina,  Greece}
\IEEEauthorblockA{\IEEEauthorrefmark{3}Department of Electrical and Computer Engineering, Aristotle University of Thessaloniki, Greece}
\IEEEauthorblockA{\IEEEauthorrefmark{4}Dienekes SI IKE, Heraklion, Greece}
\IEEEauthorblockA{\IEEEauthorrefmark{5}Department of Electrical and Computer Engineering, Technical University of Crete, Chania, Greece}
}

\maketitle


\begin{abstract}
Next-generation wireless networks are envisioned to achieve reliable, low-latency connectivity within environments characterized by strong multipath and severe channel variability. Programmable wireless environments (PWEs) address this challenge by enabling deterministic control of electromagnetic (EM) propagation through software-defined reconfigurable intelligent surfaces (RISs). However, effectively configuring RISs in real time remains a major bottleneck, particularly under near-field conditions where mutual coupling and specular reflections alter the intended response. To overcome this limitation, this paper introduces MATCH, a physics-based codebook compilation algorithm that explicitly accounts for the EM coupling among RIS unit cells and the reflective interactions with surrounding structures, ensuring that the resulting codebooks remain consistent with the physical characteristics of the environment. Finally, MATCH is evaluated under a realistic EM framework incorporating mutual coupling and secondary reflections, demonstrating its ability to concentrate scattered energy within the focal region, confirming that physics-consistent, codebook-based optimization constitutes an effective approach for practical and efficient RIS configuration.
\end{abstract}



\IEEEpeerreviewmaketitle

\section{Introduction}
Next-generation wireless networks aim to provide ultra-reliable, low-latency connectivity capable of supporting emerging services such as autonomous mobility, wireless power transfer, and extended reality. However, meeting these requirements necessitates overcoming the inherent randomness of wireless propagation, where path loss, shadowing, and multipath fading impose stringent limitations on Quality-of-Service (QoS). In this direction, Programmable Wireless Environments (PWEs) have emerged as a transformative paradigm in which the wireless medium itself becomes a programmable entity, enabling the deterministic control of wave propagation through softwared-defined components. At the core of this concept lie Reconfigurable Intelligent Surfaces (RISs), thin metamaterial-based tiles composed of two-dimensional arrays of subwavelength unit cells that integrate active components, such as PIN diodes, whose software-controlled states tune the surface currents induced by the impinging wave to produce a desired macroscopic electromagnetic (EM) response~\cite{liaskos2026tutorialcontrollingmetasurfacesnetwork}. Thus, as the effectiveness of PWEs depends on how accurately RISs are configured, developing efficient and physically consistent methodologies for RIS control becomes essential to sustain the stringent demands of next-generation wireless services in real time.

While PWEs enable the dynamic control of EM propagation, the real-time configuration requirement imposes a fundamental need for low-complexity methodologies capable of coordinating thousands of unit cells simultaneously. To alleviate this computational burden, codebook-based solutions have been introduced, where a finite set of optimized reflection patterns is precomputed and stored for rapid deployment during operation~\cite{11275068, TNSM_RIS_NETWORK}. In more detail, the codebook is compiled offline for a known and static deployment topology: for each supported functionality and candidate user/target location, an optimization loop computes the corresponding RIS reflection coefficients and stores the optimum as a codebook entry \cite{ComMag}. Since this procedure does not occur during operation, its computational load is not the limiting factor. In the research bibliography, most existing approaches have been developed under simplified far-field assumptions, which fail to capture the EM behavior that arises in the near-field regime, where mutual coupling among RIS unit cells and specular reflections from surrounding structures significantly influence the RIS response. In this direction, recent studies have investigated near-field codebook design~\cite{Suyu2024} and coupling-aware modeling~\cite{Pinjun2024}, yet these efforts generally treat the two effects in isolation or through approximate formulations. 

In this paper, we present MATCH, a physics-based codebook compilation algorithm tailored for realistic RIS-assisted PWEs. MATCH targets near-field beam focusing and bridges geometric-optics initialization with physics-aware, multi-stage refinement. It exploits the physics knowledge gathered during its execution to accelerate convergence and manages the propagated field holistically to concentrate energy within the receiver's region for enhanced performance.
The performance of MATCH is assessed through physics-based EM simulations that capture complete EM interactions, including effects critical to the macroscopic RIS response such as mutual coupling among elements and specular reflections.


\section{System Model} \label{sec:RIS_modeling}


We consider a three-dimensional indoor environment of cubic dimensions~$L_{th}$, where a transmitter located at position~\(\mathbf{s}\in[0,L_{th}]^3\) radiates a continuous-wave signal of frequency~$f$, corresponding to wavelength~$\lambda=c_0/f$ and wavenumber~$k=2\pi/\lambda$. Within this environment, one wall is coated with a RIS unit that alters the EM behavior of the space by controlling the phase and amplitude of reflected waves. In more detail, the deployed RIS is composed of $N=N_\mathrm{d}\times N_\mathrm{d}$ sub-wavelength unit cells arranged on a uniform grid with center-to-center spacing~$d$, typically satisfying $d\le\lambda/4$ to ensure accurate spatial sampling of the impinging field~\cite{9374451}. Moreover, each unit cell acts as a passive scatterer characterized by a complex reflection coefficient~$\Gamma_n=\rho_n e^{\mathrm{i}\phi_n}$, where the amplitude~$\rho_n=1$ and the phase~$\phi_n$ is electronically controlled, while its spatial characteristics are defined by the geometric center~\(\mathbf{p}_n=(x_n,y_n,z_n)\) and the outward normal vector~$\mathbf{n}_w$ of the wall on which it is mounted. 

Each unit cell functions as an individually tunable scatterer. Therefore, the field impinging on every cell needs to be evaluated explicitly, ensuring that spatial variations across the surface are faithfully represented and that the influence of the surrounding environment is properly captured. The impinging electric field upon each unit cell is shaped by multiple physical phenomena, including the direct illumination from the transmitter, the EM interactions with neighboring cells, and the secondary reflections generated by the environment. As a result, the incident field on the $n^{th}$ unit cell can be expressed as
\begin{equation}
E_{\mathrm{inc},n} = E_{\mathrm{dir},n} + E_{\mathrm{cpl},n} + \sum_{w\in\mathcal{W}} E_{\mathrm{sec},n}^{(w)},
\end{equation}
where $E_{\mathrm{dir},n}$ denotes the field component directly radiated from the transmitter toward the $n^{th}$ unit cell, $E_{\mathrm{cpl},n}$ represents the coupling arising from the interaction with neighboring cells, and $E_{\mathrm{sec},n}^{(w)}$ accounts for the secondary reflections generated by the walls, while $\mathcal{W}$ denotes the set of reflective walls within the environment. Specifically, the direct component $E_{\mathrm{dir},n}$ can be modeled as a spherical wave radiated by the transmitter, where, by setting $\mathbf{d}_n=\mathbf{p}_n-\mathbf{s}$ and $r_n=\lVert\mathbf{d}_n\rVert$, $E_{\mathrm{dir},n}$ is expressed as
\begin{equation}
E_{\mathrm{dir},n}
= \frac{e^{\mathrm{i}k r_n}}{r_n}
\underbrace{\bigl|\hat{\mathbf{d}}_n\cdot\mathbf{u}_\mathrm{TX}\bigr|^m}
_{\text{transmitter pattern}}
\underbrace{\bigl[\max\{0,(\hat{\mathbf{d}}_n\cdot\mathbf{n}_w)\}\bigr]^{p}}
_{\text{element cosine law}},
\label{eq:Edirect}
\end{equation}
where $\hat{\mathbf{d}}_n=\mathbf{d}_n/r_n$ is the unit vector pointing from the transmitter to the $n$-th unit cell, $\mathbf{u}_\mathrm{TX}$ is the transmitter’s main-lobe direction, $m$ controls the beamwidth of the radiation pattern, and $p$ controls the angular sensitivity of the unit cell, with larger values producing a more specular response and smaller values yielding a diffuse, cosine-like behavior~\cite{pitilakis2022multifunctional}. Moreover, $E_{\mathrm{cpl},n}$ quantifies the mutual coupling among neighboring cells, capturing the fact that the field scattered by one cell can re-illuminate its immediate neighbors. Thus, to describe this effect, and by taking into account a local interaction set $\mathcal{N}n$ that contains the nearest neighbors of the $n$-th unit cell, $E_{\mathrm{cpl},n}$ can be given as
\begin{equation}
E_{\mathrm{cpl},n} = \alpha \sum_{m\in\mathcal{N}_n} \Gamma_m E_{\mathrm{inc},m}
\frac{e^{\mathrm{i}k\lVert\mathbf{p}_n-\mathbf{p}_m\rVert}}
{\lVert\mathbf{p}_n-\mathbf{p}_m\rVert},
\label{eq:mutual_coupling}
\end{equation}
where $\alpha \in [0,1]$ is a dimensionless coefficient that determines the strength of the coupling interactions. Finally, to capture the additional illumination paths via specular reflections from the parts of the walls that are not covered with RIS panels, these uncovered surfaces are modeled as smooth reflectors that redirect the transmitted wave toward the RIS-coated regions. Hence, by defining $\mathbf{s}_w$ as the mirror image of the transmitter with respect to wall $w$, the secondary contribution arriving at the $n$-th unit cell from that wall can be expressed as
\begin{equation}
E_{\mathrm{sec},n}^{(w)}
= \beta_w\frac{e^{\mathrm{i}k\lVert\mathbf{p}_n-\mathbf{s}_w\rVert}}
{\lVert\mathbf{p}_n-\mathbf{s}_w\rVert}
\bigl|\hat{\mathbf{d}}_{n,w}\cdot\mathbf{n}_w\bigr|^{p},
\end{equation}
where $\hat{\mathbf{d}}_{n,w}$ is the unit vector from $\mathbf{s}_w$ to $\mathbf{p}_n$, $\mathbf{n}_w$ is the normal vector of wall $w$, and $\beta_w$ denotes its reflectivity coefficient.

Once the incident field $E_{\mathrm{inc},n}$ reaches the RIS surface, it interacts with the individual unit cells, each of which re-radiates the impinging wave after applying a controllable phase shift governed by its complex reflection coefficient~$\Gamma_n$. The collective re-radiation from all cells combines coherently in space, forming the total electric field $E(\mathbf{r})$ at any observation point~$\mathbf{r}\in[0,L_{th}]^3$, which can be expressed as
\begin{equation}
E(\mathbf{r}) = \sum_{n=1}^N \Gamma_n E_{\mathrm{inc},n}
\frac{e^{\mathrm{i}k\lVert\mathbf{r}-\mathbf{p}_n\rVert}}
{\lVert\mathbf{r}-\mathbf{p}_n\rVert}.
\label{eq:fieldSum}
\end{equation}
Through this superposition, the overall response of the PWE emerges as a function of both the incident illumination and the phase configuration imposed across the surface. However, since the resulting field inherently divides into regions of useful concentration, directed but unfocused radiation, and residual spillover, the total scattered energy can be naturally partitioned into three components corresponding to these domains. The field magnitude $E$ is therefore integrated as the squared amplitude over each region and normalized by the total scattered energy, yielding the normalized energy fractions
\begin{align}
\eta_{\mathrm{focus}} &= \frac{\sum_{\text{in}} |E|^2}{\sum_{\text{tot}} |E|^2}, \\
\eta_{\mathrm{dirOut}} &= \frac{\sum_{\text{near-out}} |E|^2}{\sum_{\text{tot}} |E|^2}, \\
\eta_{\mathrm{unexp}} &= \frac{\sum_{\text{far-out}} |E|^2}{\sum_{\text{tot}} |E|^2},
\end{align}
where $\eta_{\mathrm{focus}}$ quantifies the useful energy concentrated within the focal region, $\eta_{\mathrm{dirOut}}$ captures the directed energy that remains within the main beam but outside the focus, and $\eta_{\mathrm{unexp}}$ represents the unexploited leakage distributed throughout the environment, and $\eta_{\mathrm{focus}} + \eta_{\mathrm{dirOut}} + \eta_{\mathrm{unexp}} = 1$. Together, these quantities provide a physically complete basis for evaluating focusing efficiency and leakage suppression in PWEs. As a result, achieving accurate beam focusing requires a systematic methodology for determining the appropriate configuration of the unit cells.

We model the RIS as ideal and lossless, using unit-modulus reflection coefficients. The framework directly extends to lossy RISs by imposing an amplitude constraint $|\Gamma_n|<1$ and/or using per–unit-cell measured (or simulated) amplitude–phase responses. We further assume identical RIS units with the same coupling characteristics, and account for mutual coupling via an effective parameter that mainly captures interactions within a local neighborhood of adjacent elements. This coupling model—its strength and neighborhood definition—can be re-parameterized for different RIS designs (e.g., element geometry, inter-element spacing, substrate, biasing network) using calibration measurements or manufacturer characterization.

\section{MATCH: A Novel Physics-Based Codebook Compilation Algorithm}\label{sec:match}

In this section, we present MATCH, a physics-driven algorithm for near-field beam focusing that explicitly incorporates mutual coupling among RIS elements and specular reflections within the environment. As outlined in Alg.1, MATCH follows a multi-stage refinement process that not only derives the optimal reflection coefficients $\Gamma=\left(\Gamma_1,\dotsc,\Gamma_N\right)\in\mathbb{C}^N$for a given topology but also leverages the physics knowledge gathered during execution to improve convergence and performance. Furthermore, it formulates beam focusing as a holistic EM energy concentration problem—rather than merely maximizing energy within the focus area, it actively redirects power from undesired directions toward the region of interest.

\begin{figure}[t]
  \centering
  \includegraphics[width=\linewidth]{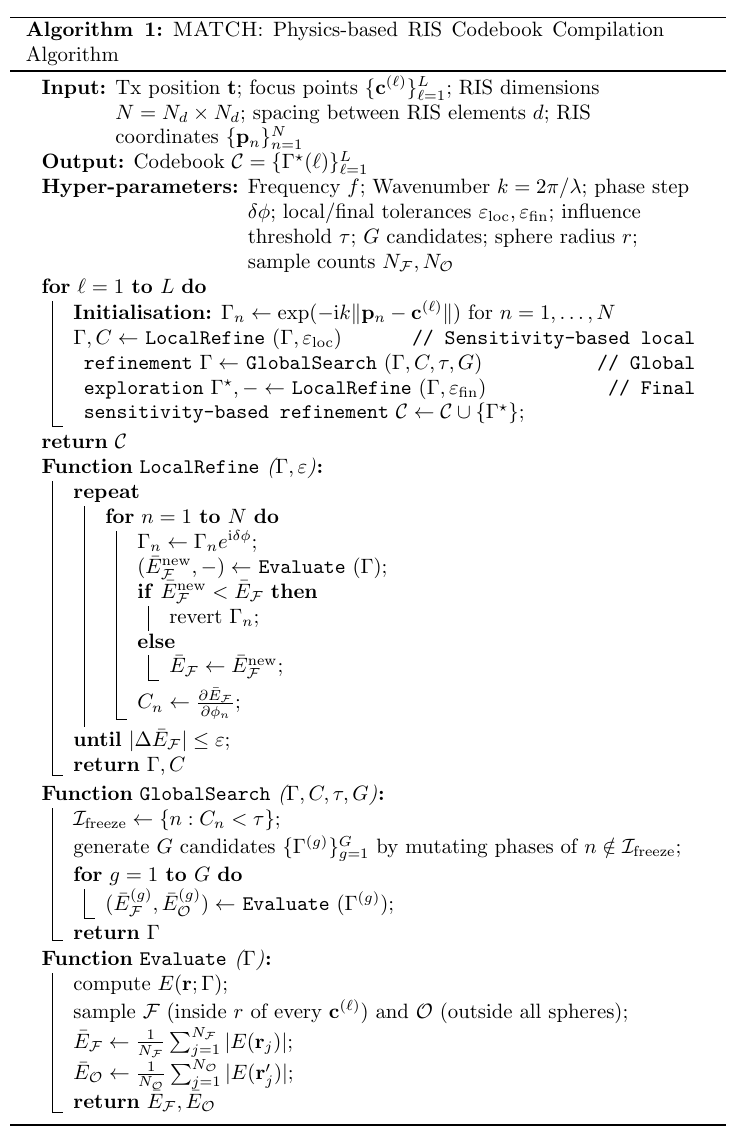}
\end{figure}

To formulate the design objective of MATCH, the desired EM behavior is defined over specific regions of the environment where field enhancement or suppression is required. The receiver area corresponds to a focus region~$\mathcal{F}$ centered at one or more points~$\{\mathbf{c}^{(\ell)}\}{\ell=1}^L$, each modeled as a small sphere of radius~$r$ that represents the spatial extent within which a strong and uniform field distribution is desired. To complement this focusing objective, a second region~$\mathcal{O}$ is defined to represent the remaining part of the environment, excluding all focus spheres, where the field intensity should be minimized. However, since the EM field distribution is inherently coupled across space, the concentration of energy within~$\mathcal{F}$ cannot be achieved independently of how the field behaves elsewhere. Consequently, an exclusive maximization of the field inside the target region may not effectively redirect the scattered energy toward it, as undesired constructive interference outside~$\mathcal{F}$ could still dominate the overall response. A formulation that simultaneously enhances the field within~$\mathcal{F}$ and attenuates it across~$\mathcal{O}$ therefore ensures that the optimization acts coherently on the global energy balance of the environment, steering the configuration toward physically consistent focusing behavior. To evaluate these objectives efficiently, the entire space is discretized into observation points at which $E(\mathbf{r})$ is computed, with a total of~$N_{\mathcal{F}}$ samples distributed within the focus regions and~$N_{\mathcal{O}}$ samples placed throughout the outer region. The sampling density is intentionally increased inside~$\mathcal{F}$ to capture fine spatial variations of the field, whereas a coarser grid is used in~$\mathcal{O}$ to maintain computational tractability. Based on this discretization, the first design goal of MATCH is to maximize the mean field magnitude $\bar{E}_{\mathcal{F}}$ within the focus regions, expressed as
\begin{equation}
\max_{\Gamma} \bar{E}_{\mathcal{F}} = \max_{\Gamma}\sum_{j=1}^{N_{\mathcal{F}}} w_j \,\bigl|E(\mathbf{r}_j)\bigr|,
\label{eq:max_field_inside}
\end{equation}
where $w_j=1/N_{\mathcal{F}}$ is a weighting factor introduced for the spatial averaging of the field over the focus region under uniform sampling, while the second goal is to minimize the corresponding mean field $\bar{E}_{\mathcal{O}}$ magnitude across the outer region, given by
\begin{equation}
\min_{\Gamma}\bar{E}_{\mathcal{O}} = \min_{\Gamma}\sum_{j=1}^{N_{\mathcal{O}}} w’_j \,\bigl|E(\mathbf{r}’_j)\bigr|,
\label{eq:min_field_outside}
\end{equation}
where $w’_j=1/N_{\mathcal{O}}$ denotes the analogous weighting coefficient associated with each sampling point in the outer region. Therefore, through the joint optimization of $\bar{E}_{\mathcal{F}}$ and $\bar{E}_{\mathcal{O}}$, MATCH establishes a physically consistent trade-off that achieves strong field focusing at the receiver while suppressing unwanted radiation throughout the rest of the environment.

Recognizing that the optimization problem addressed by MATCH is inherently nonlinear and tightly coupled to the underlying EM physics, the choice of initial conditions plays a crucial role in determining both convergence behavior and the quality of the final configuration. To ensure that the optimization starts from a physically meaningful state, MATCH employs an initialization inspired by geometric optics (GO), where each RIS element is assigned a reflection phase that promotes constructive interference at the desired focus regions (Initialization in Alg.~1. In this preliminary stage, secondary reflections and mutual coupling effects are disregarded, allowing the phase of each element to be determined solely by its propagation distance to the center of the focus region.

Following the GO initialization, MATCH transitions to a sensitivity-based local refinement stage (Stage 1) that builds directly upon the physics of the full propagation model in Eq.~\eqref{eq:fieldSum}. At this stage, mutual coupling among unit cells and secondary wall reflections are incorporated to account for the complete EM interactions that characterize the RIS-assisted channel. To consistently integrate these effects within the optimization process, each RIS element undergoes small iterative phase perturbations of magnitude~$\delta\phi$, and after each adjustment, the total field~$E(\mathbf{r})$ is recomputed across the sampling sets to evaluate the corresponding variation in the focusing objective~$\bar{E}_{\mathcal{F}}$. By continuously comparing successive field realizations, the procedure selectively retains phase updates that enhance the objective while reverting those that degrade it, thereby steering the configuration toward a physically consistent local optimum. Through this mechanism, the refinement process effectively performs a gradient-driven adaptation of the RIS phases, following the established principles of beamforming optimization, suitably extended to the near-field and strongly coupled regime of indoor RIS environments~\cite{balanis2016antenna}. The iterations proceed until the relative improvement in~$\bar{E}{\mathcal{F}}$ falls below a local tolerance~$\varepsilon_{\mathrm{loc}}$, indicating convergence to a locally stable configuration. In parallel, the same perturbation–response mechanism quantifies the contribution of each RIS element to the overall focusing performance through its sensitivity
\begin{equation}
C_n = \frac{\partial \bar{E}_{\mathcal{F}}}{\partial \phi_n},
\label{eq:sensitivity}
\end{equation}
which can be obtained analytically from the field derivative
\begin{equation}
\frac{\partial E(\mathbf{r})}{\partial \phi_n}
= j\,\Gamma_n E_{\mathrm{inc},n}\,G(\mathbf{r},\mathbf{p}_n),
\label{eq:field_derivative}
\end{equation}
where $G(\mathbf{r},\mathbf{p}_n)$ denotes the Green’s function linking the $n$-th RIS element with the observation point~$\mathbf{r}$. Aggregating this response over all focus samples yields
\begin{equation}
\frac{\partial |E|}{\partial \phi_n}
= \mathrm{Re}\!\left\{
\frac{E^*(\mathbf{r})}{|E(\mathbf{r})|}\,
\frac{\partial E(\mathbf{r})}{\partial \phi_n}
\right\},
\label{eq:field_magnitude_derivative}
\end{equation}
providing a closed-form relation between local phase adjustments and the global field strength inside~$\mathcal{F}$. 
For each sampling point~$\mathbf{r}_j$, the vector of element-wise sensitivities~$\mathbf{c}_j$ is defined with entries
\begin{equation}
[\mathbf{c}_j]_n =
\mathrm{Re}\!\left\{
\frac{E^*(\mathbf{r}_j)}{|E(\mathbf{r}_j)|}\,
j\,\Gamma_n E_{\mathrm{inc},n} G(\mathbf{r}_j,\mathbf{p}_n)
\right\},
\label{eq:sensitivity_vector}
\end{equation}
and the ensemble of these local responses is then aggregated to form the overall correlation matrix
\begin{equation}
\mathbf{H} =
\sum_{j=1}^{N_{\mathcal{F}}} \mathbf{c}_j \mathbf{c}_j^{\!\top},
\label{eq:correlation_matrix}
\end{equation}
which captures the joint influence of all RIS elements on the mean field within the focus region. The diagonal entries of~$\mathbf{H}$ quantify the individual contribution of each element to the focusing performance, whereas the off-diagonal terms characterize inter-element coupling due to mutual interactions and shared propagation paths. Consequently, $\mathbf{H}$ provides a physically grounded foundation for identifying low-influence unit cells that can be frozen to improve efficiency and for defining high-impact unit cells that can guide a global optimization phase.

After reaching local convergence, MATCH naturally transitions to a global exploration phase (Stage 2) that extends the optimization beyond the neighborhood shaped by the sensitivity-based refinement. Building upon ~$C_n^{(t)}$, this stage seeks to reconcile the two physically intertwined objectives of enhancing the field intensity within the focus region~$\bar{E}_{\mathcal{F}}$ while simultaneously suppressing the residual radiation across the outer region~$\bar{E}_{\mathcal{O}}$. However, since constructive interference and leakage suppression are inherently coupled EM phenomena, their joint control  requires a Pareto-based optimization framework that captures the fundamental trade-off between them. Within this setting, the sensitivity data guide the search process by identifying elements with limited impact that can be temporarily frozen, thereby constraining the exploration to the most influential degrees of freedom without compromising physical accuracy. This targeted reduction of the search space enables the algorithm to concentrate computational resources on the most influential unit cells and efficiently traverse the Pareto landscape, where each candidate configuration is evaluated through the pair of metrics $(\bar{E}_{\mathcal{F}},\bar{E}_{\mathcal{O}})$. Finally, to balance convergence efficiency and diversity among solutions, MATCH employs a class of multi-objective metaheuristics, with NSGA-II being particularly effective due to its ability to preserve Pareto front diversity and achieve rapid convergence in physics-driven optimization scenarios~\cite{ComMag}. Once a balance between focusing gain and leakage suppression is reached, a final sensitivity-based refinement (Stage 3) is performed through gradient ascent to fine-tune all RIS phases under the complete propagation model, and the resulting configuration~$\Gamma^\star$ is stored in the codebook, completing one MATCH cycle.

\section{Simulation Results} \label{sec:evaluation}

\begin{figure*}[t]
  \centering
  \begin{subfigure}[b]{0.3\textwidth}
    \includegraphics[width=\linewidth]{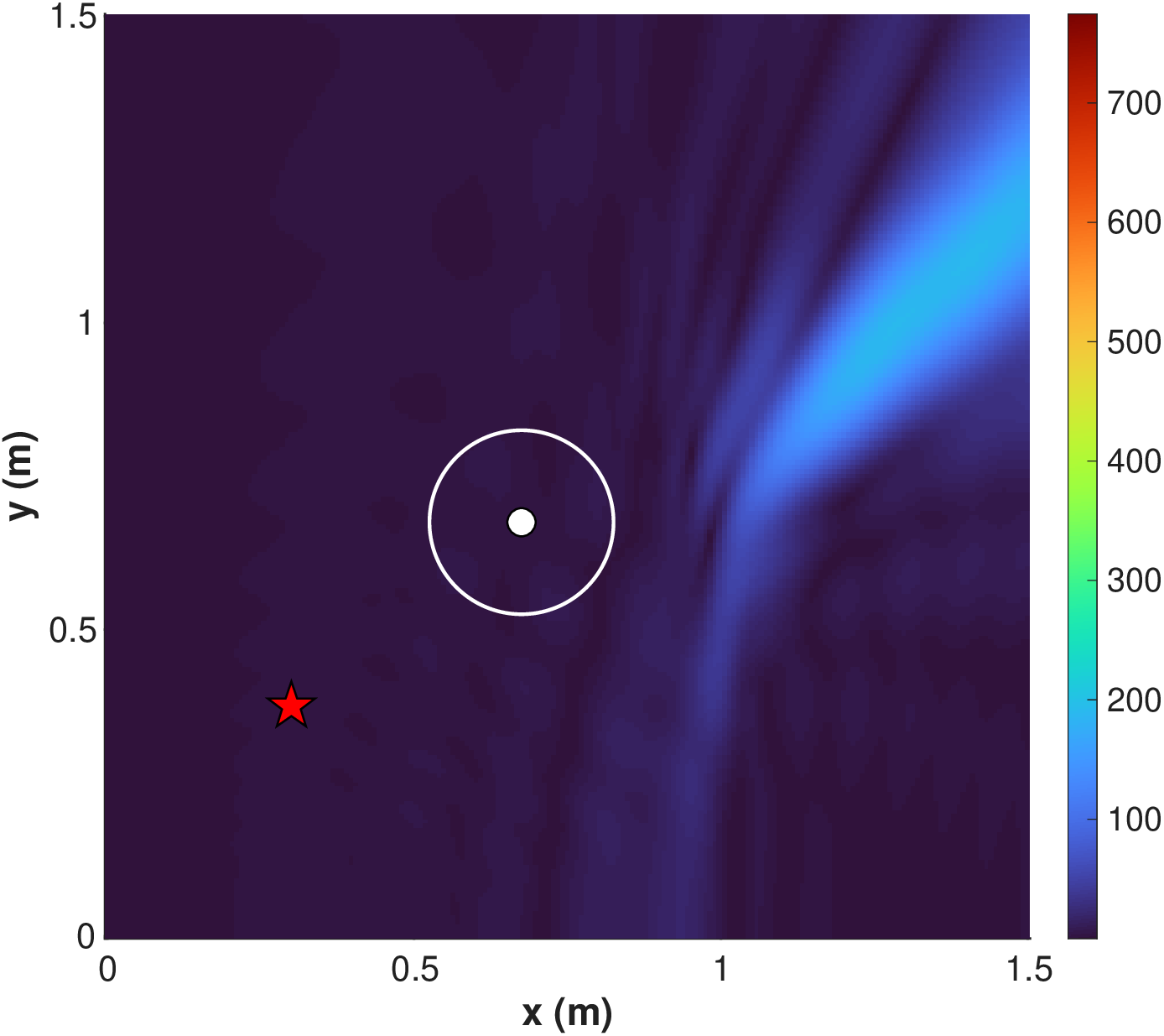}
    \caption{}
    \label{fig:OneRIS_GO}
  \end{subfigure}\hfill%
  \begin{subfigure}[b]{0.3\textwidth}
    \includegraphics[width=\linewidth]{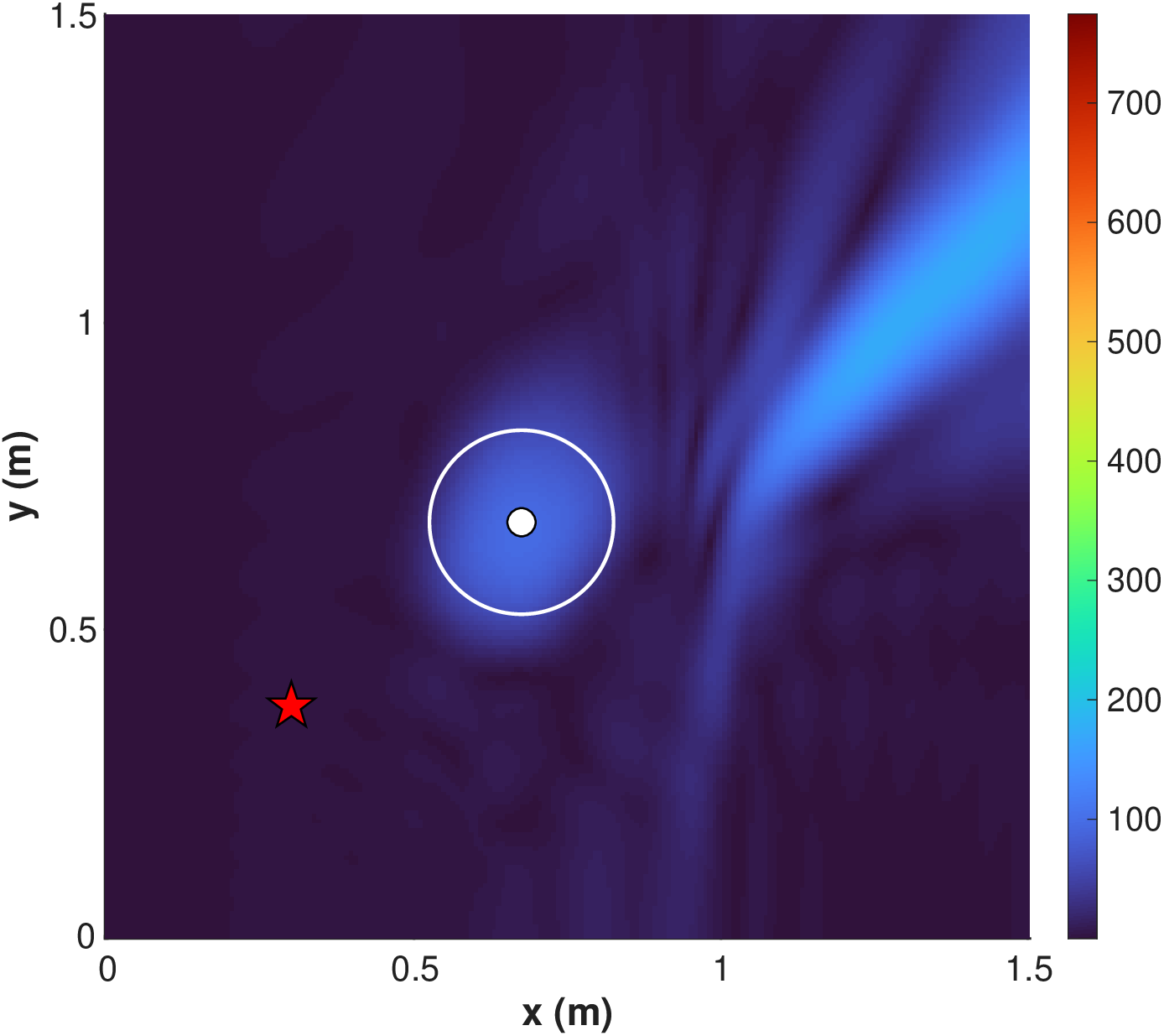}
    \caption{}
    \label{fig:OneRIS_GRAD}
  \end{subfigure}\hfill%
  \begin{subfigure}[b]{0.3\textwidth}
    \includegraphics[width=\linewidth]{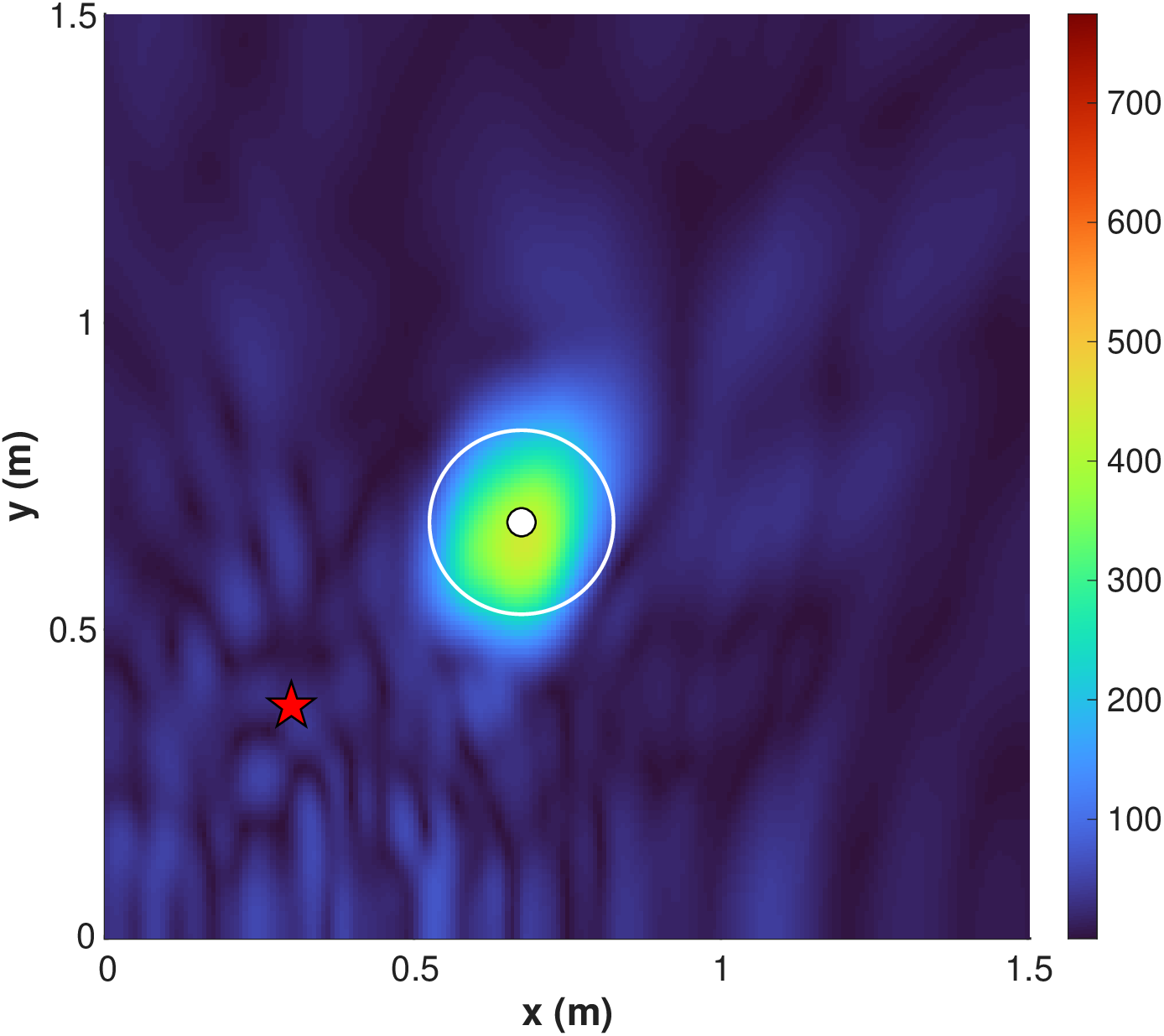}
    \caption{}
    \label{fig:OneRIS_NSGA_MATCH}
  \end{subfigure}\hfill%

  \begin{subfigure}[b]{0.3\textwidth}
    \includegraphics[width=\linewidth]{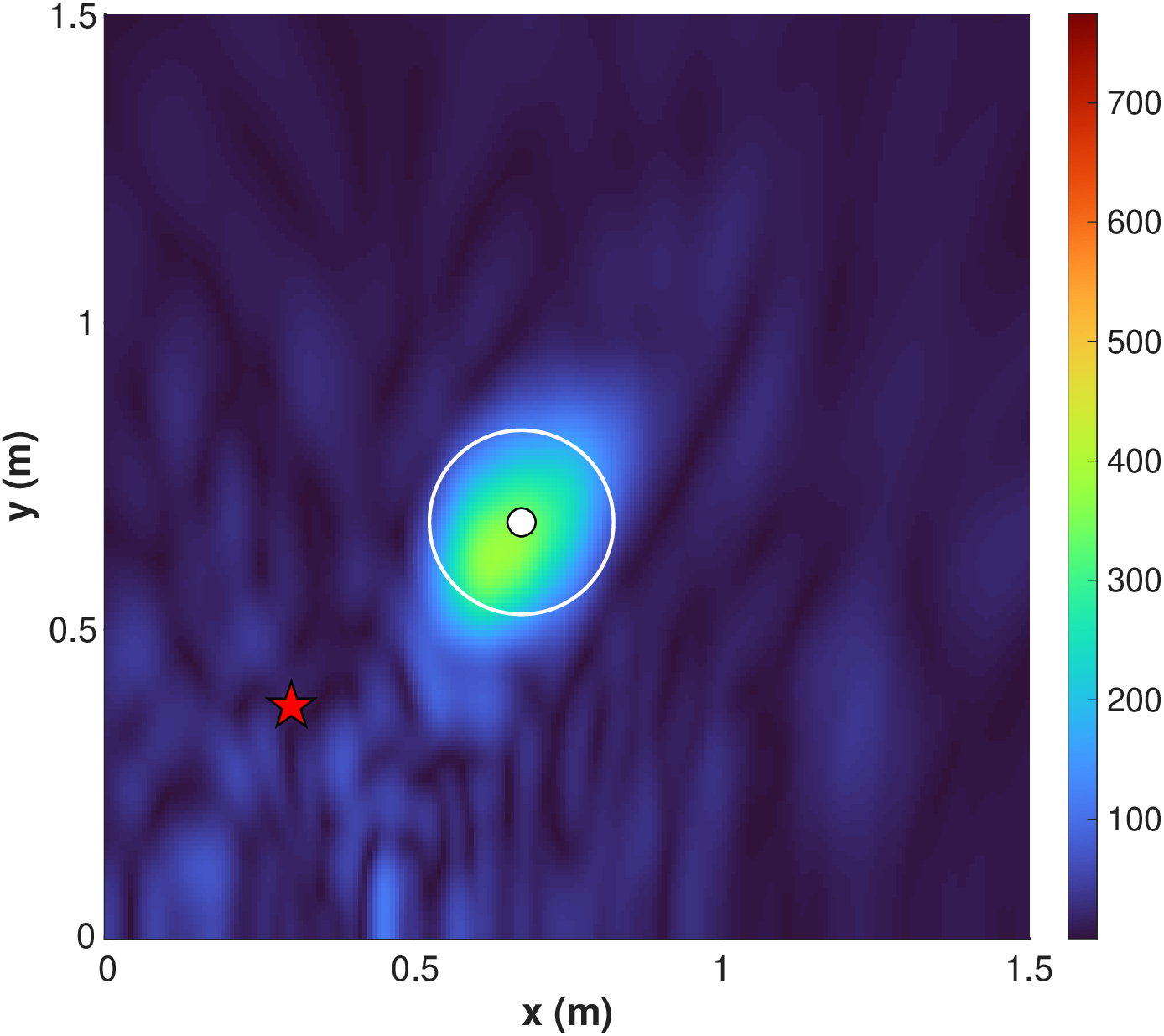}
    \caption{}
    \label{fig:OneRIS_NSGA_withoutMATCH}
  \end{subfigure}\hfill%
  \begin{subfigure}[b]{0.3\textwidth}
    \includegraphics[width=\linewidth]{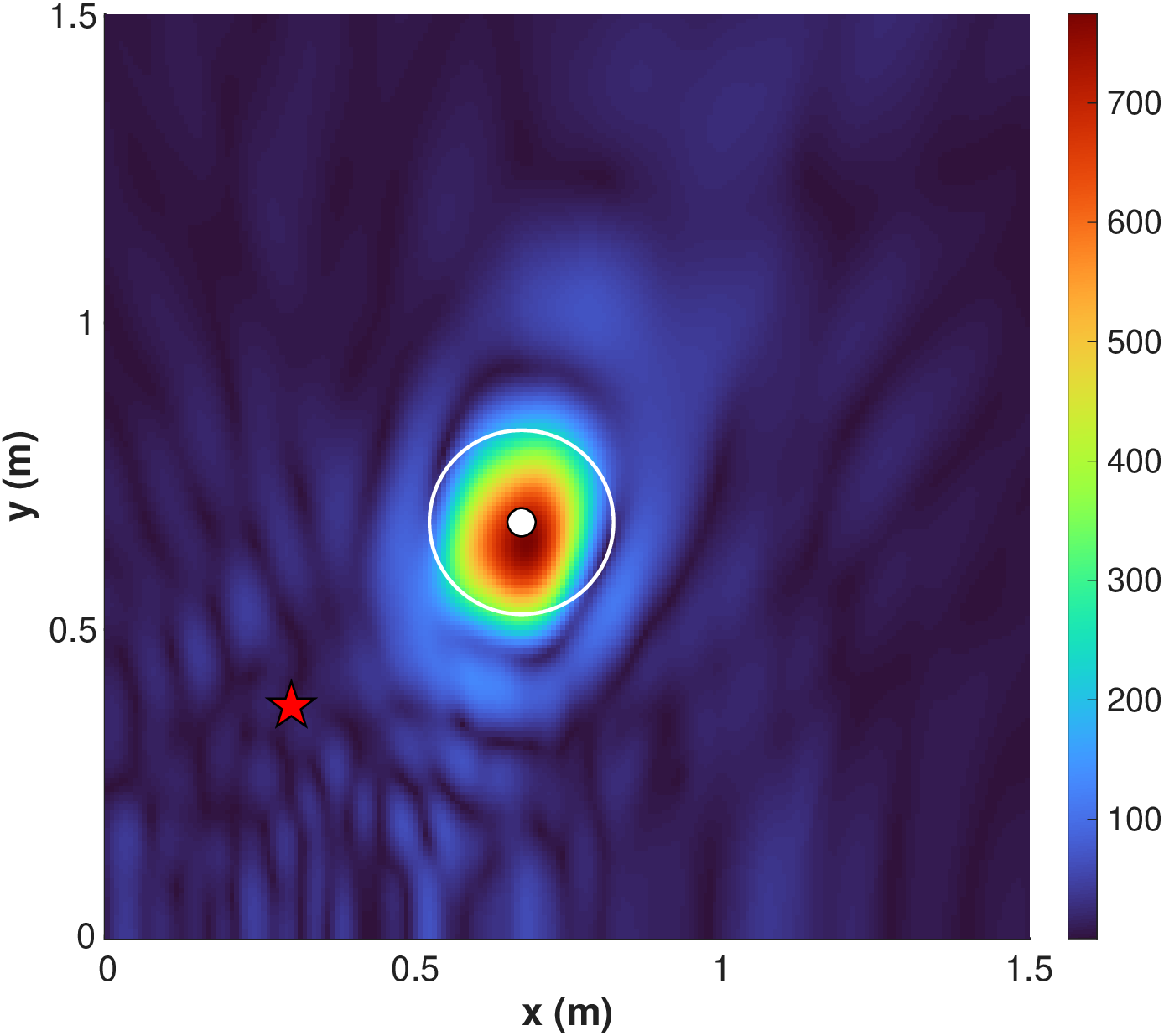}
    \caption{}
    \label{fig:OneRIS_final_MATCH}
  \end{subfigure}\hfill%
  \begin{subfigure}[b]{0.3\textwidth}
    \includegraphics[width=\linewidth]{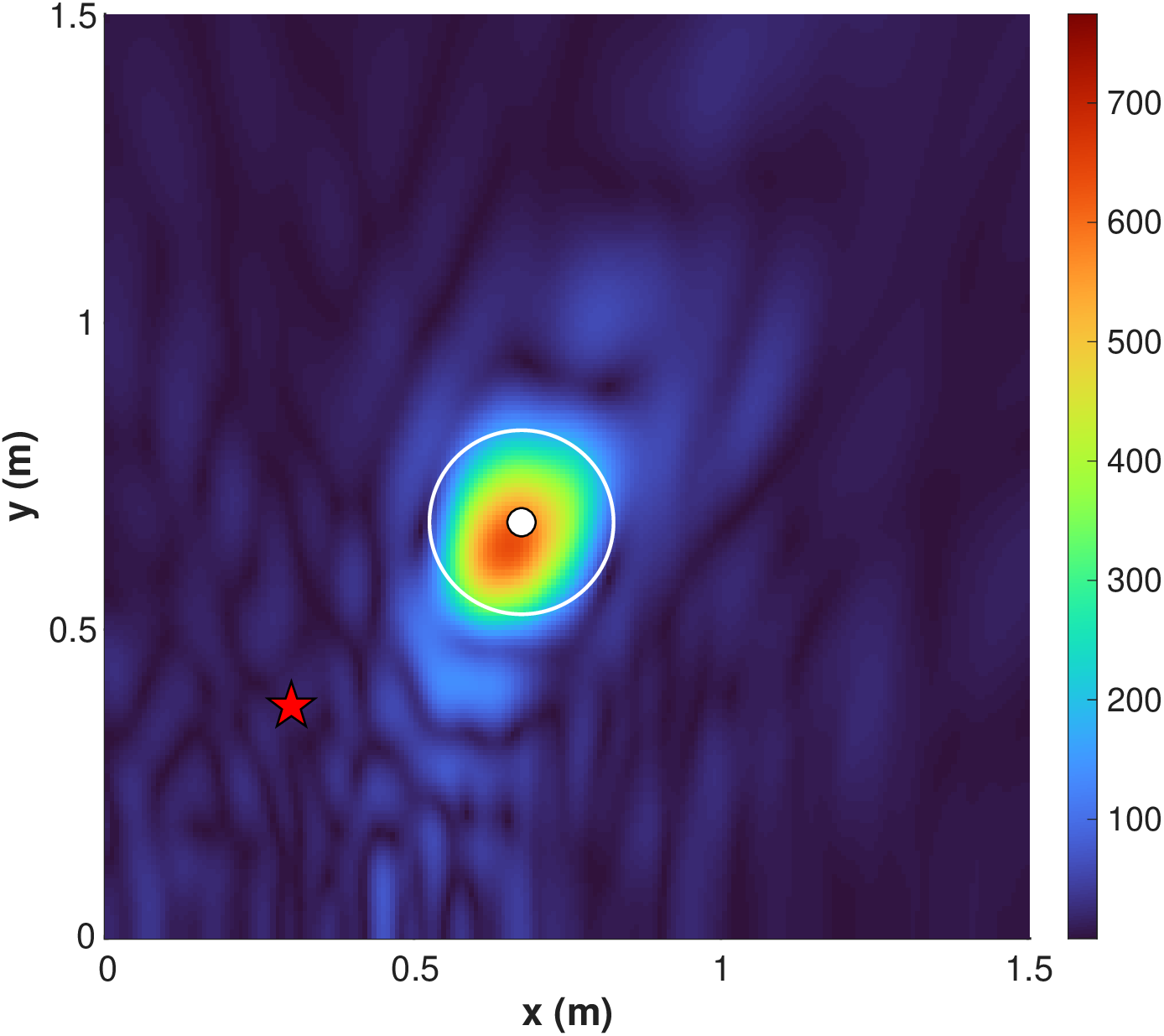}
    \caption{}
    \label{fig:OneRIS_final_withoutMATCH}
  \end{subfigure}\hfill%

  \caption{Near-field beam-focusing patterns: (a); (b) Stage 1; (c,d) Stage 2 with/without minimization; (e,f) Stage 3 with/without minimization. Red star = transmitter; white circle = Receiver /focus area.}
  \label{fig:results_oneRIS}
\end{figure*}

In this section, we evaluate the performance of MATCH. The corresponding physical and algorithmic parameters used throughout all experiments are summarized in Table~\ref{table1}. The initialization of the phase perturbation step is set to $\delta\phi = 0.2$ with a local convergence tolerance of $\varepsilon_{\mathrm{loc}} = 10^{-2}$, ensuring a balanced trade-off between exploration and stability during the early optimization stages, while in the subsequent refinement iterations these values are reduced to $\delta\phi = 0.1$ and $\varepsilon_{\mathrm{fin}} = 10^{-5}$ to enable smooth convergence toward the final optimum without oscillatory behavior. Moreover, each optimization run is restricted to a maximum of $10^4$ iterations, providing a consistent balance between computational efficiency and accuracy across all configurations. Furthermore, the perturbation–response information obtained during the sensitivity analysis is leveraged to gradually focus the optimization on the most relevant degrees of freedom by freezing the least-contributing $5\%$ of RIS elements every $25\%$ of the total generations, thereby allocating computational effort to the most influential unit cells, enabling faster convergence and improved field control. Finally, after convergence, the resulting field distributions are evaluated on a horizontal plane intersecting the receiver level, and all radiation patterns are normalized with respect to an incident field magnitude of $E_{\mathrm{n}} = 1$ V/m  to enable consistent visual and quantitative comparisons across different setups.

\begin{table}[t]
\footnotesize
\renewcommand{\arraystretch}{1.1}
\caption{\textsc{Simulation Parameters}}
\label{table1}
\centering
\begin{tabular}{@{\,}ll@{\,}}
\toprule
\textbf{Parameter} & \textbf{Value} \\
\midrule
Room dimensions & $L_{\mathrm{th}} = 1.5$ m \\
RIS dimensions & $120 \times 120$ unit cells \\
Inter-element spacing & $d = \lambda / 4$ \\
Operating frequency & $f = 6$ GHz \\
Focus region radius & $r = 0.15$ m \\
Focus sampling points & $N_{\mathcal{F}} = 12{,}500$ \\
Outer sampling points & $N_{\mathcal{O}} = 15{,}000$ \\
Mutual coupling coefficient & $\alpha = 0.15$ \\
Field normalization reference & $E_{\mathrm{n}} = 1$ V/m \\
Population size & 300 \\
Generations & 75 \\
Crossover / mutation indices & 8 / 8 \\
Mutation probability & 25\% \\
\bottomrule
\end{tabular}
\end{table}

Fig.~\ref{fig:results_oneRIS} illustrates the evolution of the near-field beam focusing process throughout the stages of the MATCH algorithm for the single-RIS configuration. The sequence begins in Fig.~\ref{fig:OneRIS_GO}, where the GO initialization establishes a phase distribution that directs the reflected energy toward the intended region, forming a coherent baseline for the subsequent refinement. However, as this stage does not account for mutual coupling and specular reflections, the resulting field still exhibits significant leakage along the main beam, and the energy concentration within the focal region remains limited. When these EM interactions are incorporated through the sensitivity-based local refinement shown in Fig.~\ref{fig:OneRIS_GRAD}, the reflected wavefront becomes progressively reorganized within the space, with the RIS adapting its response in a manner consistent with mutual coupling and specular reflections. As a result, the energy density within the focus increases by $22.9$~dB, while the off-axis energy decreases, though residual sidelobes remain due to the optimization acting only on the inner region. This imbalance naturally leads to the global exploration stage illustrated in Fig.~\ref{fig:OneRIS_NSGA_MATCH}, where MATCH simultaneously maximizes the field within the focus area and minimizes it outside of the focus area, achieving an additional $11.3$~dB improvement and a noticeably sharper confinement of energy around the receiver. However, Fig.~\ref{fig:OneRIS_NSGA_withoutMATCH} highlights that when this outer-field minimization is omitted, the attainable improvement drops to $7.99$~dB, confirming that suppressing leakage indirectly reinforces the focal intensity. Finally, as the optimization converges, the configuration shown in Fig.~\ref{fig:OneRIS_final_MATCH} fine-tunes the global solution through the final sensitivity-based refinement, adding another $3.5$~dB gain and stabilizing the beam without reintroducing sidelobes. In contrast, when the same refinement follows the single-objective formulation, as in Fig.~\ref{fig:OneRIS_final_withoutMATCH}, the gain reaches only $2.5$~dB, showing that the coupled optimization is essential for accurate near-field beam focusing.

\begin{table}[ht]
\centering
\caption{Energy fractions by region for each stage of MATCH w/ \& w/o the minimization objective}
\begin{tabular}{l|ccc}
\toprule
\textbf{Stage} & $\eta_{\mathrm{focus}}$ & $\eta_{\mathrm{dirOut}}$ & $\eta_{\mathrm{unexp}}$ \\
\midrule
GO (w \& w/o)           & 0.1\%   & 6.8\%   & 93.1\% \\
\midrule
Stage 1 (w \& w/o)            & 12.62\% & 10.36\% & 77.02\% \\
\midrule
Stage 2 (w)            & 75.5\%  & 16.36\% & 8.14\% \\
Stage 2 (w/o)               & 71.75\% & 18.19\% & 10.06\% \\
\midrule
Stage 3 (w)            & 85.66\% & 11.62\% & 2.72\% \\
Stage 3 (w/o)               & 80.5\%  & 14.56\% & 4.94\% \\
\bottomrule
\end{tabular}
\label{table:percentage_results}
\end{table}

Table~\ref{table:percentage_results} complements these observations by quantifying the redistribution of the total scattered energy across the focused, directed, and unexploited components at each stage. Starting from the geometric-optics initialization, where only 0.1\% of the total energy lies within the focal region, the sensitivity-based refinement already redirects 12.6\% toward the target and reduces the leakage fraction from 93.1\% to 77.0\%, indicating that coupling and reflections begin to cooperate constructively. Building on this, the global exploration further aligns the scattered energy, raising the focused fraction to 75.5\% while lowering the unexploited energy below 10\%. The final refinement consolidates this effect, achieving 85.7\% of the total energy within the focus and reducing the remaining directed and unexploited components to 11.6\% and 2.7\%, respectively. Compared to the single-objective case, which concentrates only 80.5\% and exhibits higher leakage, these results show that effective near-field beam focusing arises not merely from constructive interference but from the coordinated adjustment of all interactions within the environment. In this way, MATCH provides not only enhanced focusing performance but also physical insight into how mutual coupling and specular reflections shape the spatial energy distribution in RIS-assisted networks.

\section{Conclusion}\label{sec:conclusion}
In this paper, we introduced MATCH, a physics-based codebook compilation algorithm for RIS-assisted programmable wireless environments that achieves accurate near-field beam focusing by jointly accounting for mutual coupling and specular reflections. MATCH progresses from a geometric-optics initialization that establishes a coherent phase baseline, to a sensitivity-based refinement that incorporates full EM interactions, followed by a global exploration guided by element influence to target high-impact regions, and a final sensitivity-based refinement that consolidates convergence and stabilizes the near-field focus. Overall, MATCH exploits the physics knowledge gathered during its execution to accelerate convergence and manages the propagated field holistically, concentrating energy within the receiver’s region for enhanced performance. These results confirm that a physics-consistent, codebook-based formulation provides an effective and interpretable path for RIS optimization, linking practical focusing performance with clear physical insight into near-field energy redistribution.
\vspace{-2mm}
\section*{Acknowledgments}
This work has been funded by the Smart Networks and Services Joint Undertaking (SNS JU) under the European Union’s Horizon Europe research and innovation programme through the 6G Trans-Continental Edge Learning (6G-XCEL) project (Grant Agreement No 101139194), and through the NATWORK project (Net-Zero self-adaptive activation of distributed self-resilient augmented services; Grant Agreement No 101139285). Within 6G-XCEL, we established the initial technical foundations and performed the final synthesis and interpretation of results, while, within NATWORK, we developed the experimental workflow and carried out the main evaluation/validation.
\bibliographystyle{ieeetr}
\bibliography{refs}

\end{document}